\pdfoutput=1

\documentclass[10pt,letter,prl,twocolumn,hypertext]{revtex4-1}

\usepackage{ifthen} 
\newboolean{pdflatex}
\setboolean{pdflatex}{true} 

\newboolean{articletitles}
\setboolean{articletitles}{true} 

\newboolean{uprightparticles}
\setboolean{uprightparticles}{false} 

\usepackage{microtype}
\usepackage{lineno}  
\usepackage{xspace} 
\usepackage{bigstrut}

\usepackage{graphicx}  
\usepackage{color}
\usepackage{colortbl}
\graphicspath{{./figs/}} 

\usepackage{amsmath} 
\usepackage{amssymb}
\usepackage{amsfonts}
\usepackage{upgreek} 

\newcommand*\patchAmsMathEnvironmentForLineno[1]{%
\expandafter\let\csname old#1\expandafter\endcsname\csname #1\endcsname
\expandafter\let\csname oldend#1\expandafter\endcsname\csname
end#1\endcsname
 \renewenvironment{#1}%
   {\linenomath\csname old#1\endcsname}%
   {\csname oldend#1\endcsname\endlinenomath}%
}
\newcommand*\patchBothAmsMathEnvironmentsForLineno[1]{%
  \patchAmsMathEnvironmentForLineno{#1}%
  \patchAmsMathEnvironmentForLineno{#1*}%
}
\AtBeginDocument{%
\patchBothAmsMathEnvironmentsForLineno{equation}%
\patchBothAmsMathEnvironmentsForLineno{align}%
\patchBothAmsMathEnvironmentsForLineno{flalign}%
\patchBothAmsMathEnvironmentsForLineno{alignat}%
\patchBothAmsMathEnvironmentsForLineno{gather}%
\patchBothAmsMathEnvironmentsForLineno{multline}%
}

\usepackage{hyperref}    
\usepackage[all]{hypcap} 

\def\ux85 {\mbox{UX85}\xspace}

\ifthenelse{\boolean{uprightparticles}}%
{

 \def\PDelta      {\ensuremath{\Delta}\xspace}                 
 \def\PXi      {\ensuremath{\Xi}\xspace}                 
 \def\PLambda      {\ensuremath{\Lambda}\xspace}                 
 \def\PSigma      {\ensuremath{\Sigma}\xspace}                 
 \def\POmega      {\ensuremath{\Omega}\xspace}                 
 \def\PUpsilon      {\ensuremath{\Upsilon}\xspace}                 
 
 \def\PB      {\ensuremath{\mathrm{B}}\xspace}                 
                  
 \def\PD      {\ensuremath{\mathrm{D}}\xspace}

 \def\PK      {\ensuremath{\mathrm{K}}\xspace}

 \def\Pi      {\ensuremath{\mathrm{i}}\xspace}

 \def\Ps      {\ensuremath{\mathrm{s}}\xspace}

}
{

 \mathchardef\PDelta="7101
 \mathchardef\PXi="7104
 \mathchardef\PLambda="7103
 \mathchardef\PSigma="7106
 \mathchardef\POmega="710A
 \mathchardef\PUpsilon="7107
                  
 \def\PB      {\ensuremath{B}\xspace}                 
                  
 \def\PD      {\ensuremath{D}\xspace}

 \def\PK      {\ensuremath{K}\xspace}

 \def\Pi      {\ensuremath{i}\xspace}

 \def\Ps      {\ensuremath{s}\xspace}

}

\def\squark    {\ensuremath{\Ps}\xspace}

\def\kaon  {\ensuremath{\PK}\xspace}
  \def\Kbar  {\kern 0.2em\overline{\kern -0.2em \PK}{}\xspace}

\def\Kz    {\ensuremath{\kaon^0}\xspace}
\def\Kzb   {\ensuremath{\Kbar^0}\xspace}
\def\KzKzb {\ensuremath{\Kz \kern -0.16em \Kzb}\xspace}
\def\Kp    {\ensuremath{\kaon^+}\xspace}
\def\Km    {\ensuremath{\kaon^-}\xspace}

\def\KpKm  {\ensuremath{\Kp \kern -0.16em \Km}\xspace}

  \def\Dbar    {\kern 0.2em\overline{\kern -0.2em \PD}{}\xspace}
\def\D       {\ensuremath{\PD}\xspace}

\def\Dz      {\ensuremath{\D^0}\xspace}
\def\Dzb     {\ensuremath{\Dbar^0}\xspace}
\def\DzDzb   {\ensuremath{\Dz {\kern -0.16em \Dzb}}\xspace}
\def\Dp      {\ensuremath{\D^+}\xspace}
\def\Dm      {\ensuremath{\D^-}\xspace}

\def\DpDm    {\ensuremath{\Dp {\kern -0.16em \Dm}}\xspace}

\def\B       {\ensuremath{\PB}\xspace}
\def\Bbar    {\ensuremath{\kern 0.18em\overline{\kern -0.18em \PB}{}}\xspace}

\def\Bd      {\ensuremath{\B^0}\xspace}
\def\Bs      {\ensuremath{\B^0_\squark}\xspace}

  \def\Y#1S{\ensuremath{\PUpsilon{(#1S)}}\xspace}

\def\L {\ensuremath{\PLambda}\xspace}
\def\Lbar {\ensuremath{\kern 0.1em\overline{\kern -0.1em\PLambda}}\xspace}

\def\to                 {\ensuremath{\rightarrow}\xspace}

\def\CP                {\ensuremath{C\!P}\xspace}

\def\FL       {\ensuremath{F_{\mathrm{L}}}\xspace}
\def\AT#1     {\ensuremath{A_{\mathrm{T}}^{#1}}\xspace}           

\def\C#1      {\ensuremath{\mathcal{C}_{#1}}\xspace}                       
\def\Cp#1     {\ensuremath{\mathcal{C}_{#1}^{'}}\xspace}                    
\def\Ceff#1   {\ensuremath{\mathcal{C}_{#1}^{\mathrm{(eff)}}}\xspace}        
\def\Cpeff#1  {\ensuremath{\mathcal{C}_{#1}^{'\mathrm{(eff)}}}\xspace}       
\def\Ope#1    {\ensuremath{\mathcal{O}_{#1}}\xspace}                       
\def\Opep#1   {\ensuremath{\mathcal{O}_{#1}^{'}}\xspace}                    

\newcommand{\tev}{\ensuremath{\mathrm{\,Te\kern -0.1em V}}\xspace}
\newcommand{\gev}{\ensuremath{\mathrm{\,Ge\kern -0.1em V}}\xspace}
\newcommand{\mev}{\ensuremath{\mathrm{\,Me\kern -0.1em V}}\xspace}
\newcommand{\kev}{\ensuremath{\mathrm{\,ke\kern -0.1em V}}\xspace}
\newcommand{\ev}{\ensuremath{\mathrm{\,e\kern -0.1em V}}\xspace}
\newcommand{\gevc}{\ensuremath{{\mathrm{\,Ge\kern -0.1em V\!/}c}}\xspace}
\newcommand{\mevc}{\ensuremath{{\mathrm{\,Me\kern -0.1em V\!/}c}}\xspace}
\newcommand{\gevcc}{\ensuremath{{\mathrm{\,Ge\kern -0.1em V\!/}c^2}}\xspace}
\newcommand{\gevgevcccc}{\ensuremath{{\mathrm{\,Ge\kern -0.1em V^2\!/}c^4}}\xspace}
\newcommand{\mevcc}{\ensuremath{{\mathrm{\,Me\kern -0.1em V\!/}c^2}}\xspace}

\def\ps   {\ensuremath{{\rm \,ps}}\xspace}

\def\gsim{{~\raise.15em\hbox{$>$}\kern-.85em
          \lower.35em\hbox{$\sim$}~}\xspace}
\def\lsim{{~\raise.15em\hbox{$<$}\kern-.85em
          \lower.35em\hbox{$\sim$}~}\xspace}

\def\photos     {\mbox{\textsc{Photos}}\xspace}

\def\tell1  {TELL1\xspace}
\def\ukl1   {UKL1\xspace}

\usepackage{mciteplus}

\usepackage{longtable} 

\begin{document}

\title{First observation of {\boldmath $C\!P$} violation in the decays of {\boldmath $B^0_s$} mesons}
\vspace*{1cm}
\author{
\begin{center}
\small
R.~Aaij$^{40}$, 
C.~Abellan~Beteta$^{35,n}$, 
B.~Adeva$^{36}$, 
M.~Adinolfi$^{45}$, 
C.~Adrover$^{6}$, 
A.~Affolder$^{51}$, 
Z.~Ajaltouni$^{5}$, 
J.~Albrecht$^{9}$, 
F.~Alessio$^{37}$, 
M.~Alexander$^{50}$
% ...

{\footnotesize \it
% ...
$ ^{5}$Clermont Universit\'{e}, Universit\'{e} Blaise Pascal, CNRS/IN2P3, LPC, Clermont-Ferrand, France\\
$ ^{6}$CPPM, Aix-Marseille Universit\'{e}, CNRS/IN2P3, Marseille, France\\
% ...
$ ^{9}$Fakult\"{a}t Physik, Technische Universit\"{a}t Dortmund, Dortmund, Germany\\
% ...
$ ^{35}$Universitat de Barcelona, Barcelona, Spain\\
$ ^{36}$Universidad de Santiago de Compostela, Santiago de Compostela, Spain\\
$ ^{37}$European Organization for Nuclear Research (CERN), Geneva, Switzerland\\
% ...
$ ^{40}$Nikhef National Institute for Subatomic Physics, Amsterdam, The Netherlands\\
% ...
$ ^{45}$H.H. Wills Physics Laboratory, University of Bristol, Bristol, United Kingdom\\
% ...
$ ^{50}$School of Physics and Astronomy, University of Glasgow, Glasgow, United Kingdom\\
$ ^{51}$Oliver Lodge Laboratory, University of Liverpool, Liverpool, United Kingdom\\
% ...
$ ^{n}$LIFAELS, La Salle, Universitat Ramon Llull, Barcelona, Spain\\
% ...
}
\end{center}
}
\collaboration{The LHCb collaboration\vspace{0.4cm}}
\begin{abstract}
Using $pp$ collision data, corresponding to an integrated luminosity of $1.0~\mathrm{fb}^{-1}$, collected by LHCb in 2011 at a center-of-mass energy of $7\!$~\tev, we report the measurement of direct \CP violation in $B^0_s \to K^-\pi^+$ decays, $A_{C\!P}(B^0_s \rightarrow K^- \pi^+)=0.27 \pm 0.04\,\mathrm{(stat)} \pm 0.01\,\mathrm{(syst)}$, with significance exceeding five standard deviations.
This is the first observation of \CP violation in the decays of $B^0_s$ mesons. Furthermore, we provide an improved determination of direct $C\!P$ violation in $B^0 \to K^+\pi^-$ decays, $A_{C\!P}(B^0\rightarrow K^+ \pi^-)=-0.080 \pm 0.007\,\mathrm{(stat)} \pm 0.003\,\mathrm{(syst)}$, which is the most precise measurement of this quantity to date.
\end{abstract}

\pacs{Valid PACS appear here}




\vspace*{-1cm}
\hspace{-8.6cm}
\mbox{\Large EUROPEAN ORGANIZATION FOR NUCLEAR RESEARCH (CERN)}

\vspace*{0.2cm}
\hspace*{-9cm}
\begin{tabular*}{16.6cm}{lc@{\extracolsep{\fill}}r}
\ifthenelse{\boolean{pdflatex}}
{\vspace*{-3.2cm}\mbox{\!\!\!\includegraphics[width=.14\textwidth]{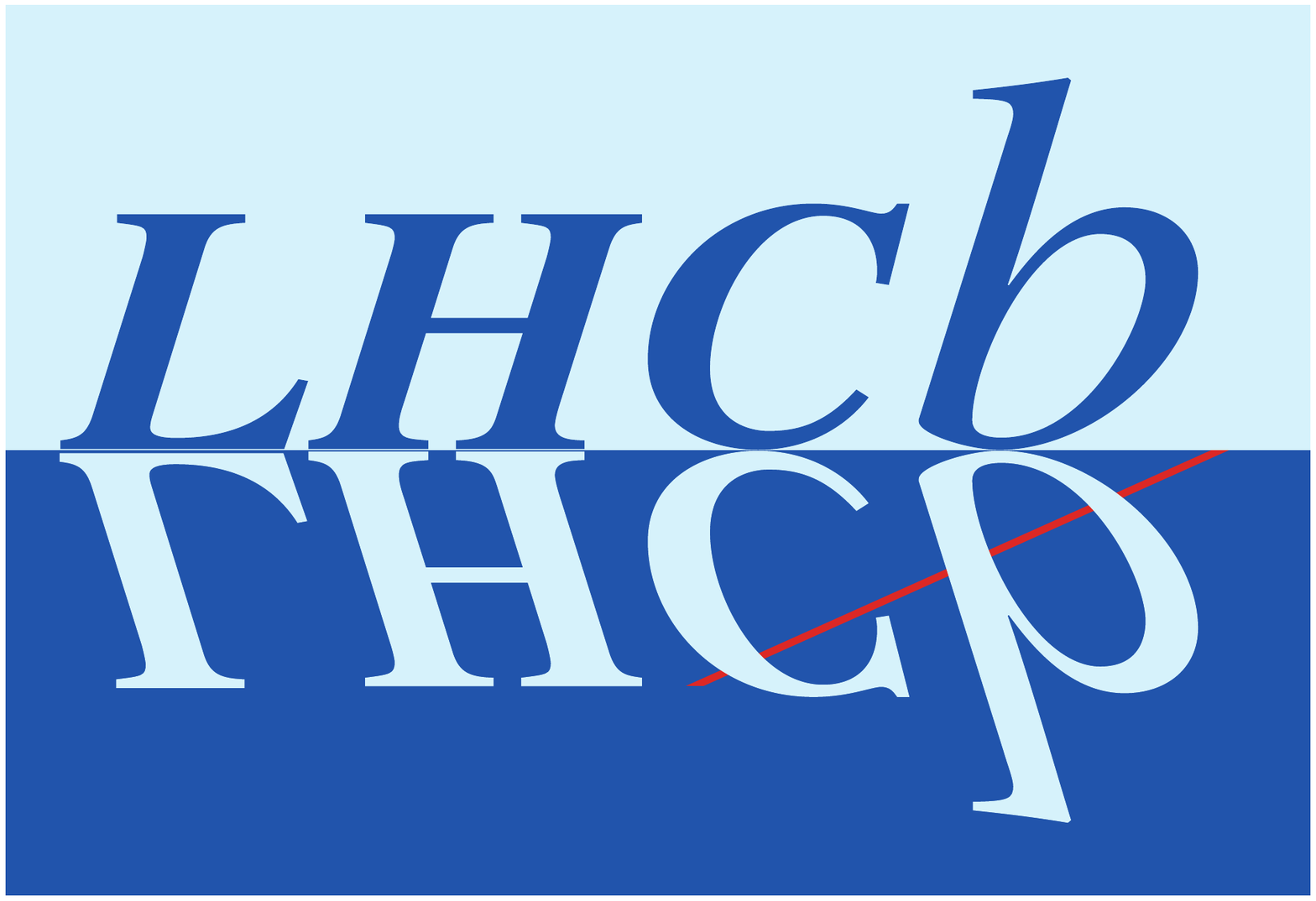}} & &}%
{\vspace*{-1.2cm}\mbox{\!\!\!\includegraphics[width=.12\textwidth]{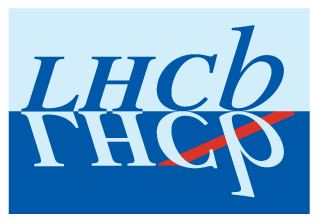}} & &}%
\\
 & & CERN-PH-EP-2013-068 \\  
 & & LHCb-PAPER-2013-018 \\  
 & & June 7, 2013 \\ 
 & & \\
\end{tabular*}

\vspace*{21.0cm}
\hspace*{-5cm}\centerline{\copyright~CERN on behalf of the LHCb collaboration, license \href{http://creativecommons.org/licenses/by/3.0/}{CC-BY-3.0}.}
\vspace*{-21.0cm}

\vspace*{20.0cm}
\hspace*{-5cm}\centerline{\large Submitted to Phys. Rev. Lett.}
\vspace*{-20.0cm}














\maketitle






%

The non-invariance of fundamental interactions under the combined action of the charge conjugation ($C$) and parity ($P$) transformations is experimentally well established in the $K^0$ and $B^0$ meson systems~\cite{Christenson:1964fg,Aubert:2001nu,Abe:2001xe,PDG2012}.
The Standard Model (SM) description of $C\!P$ violation, as given by the Cabibbo-Kobayashi-Maskawa (CKM) theory of quark-flavor mixing~\cite{Cabibbo:1963yz,Kobayashi:1973fv}, has been very successful in describing existing data. However, the source of $C\!P$ violation in the SM is known to be too small to account for the matter-dominated universe~\cite{Cohen:1993nk,Riotto:1999yt,Hou:2008xd}.

The study of $C\!P$ violation in charmless charged two-body decays of neutral $B$ mesons provides stringent tests of the CKM picture in the SM, and is a sensitive probe to search for the presence of non-SM physics~\cite{Deshpande:1994ii,He:1998rq,Fleischer:1999pa,Gronau:2000md,Lipkin:2005pb,Fleischer:2007hj,Fleischer:2010ib}. However, quantitative SM predictions for \CP violation in these decays are challenging because of the presence of hadronic factors in the decay amplitudes, which cannot be accurately calculated from quantum chromodynamics (QCD) at present. It is crucial to combine several measurements from such two-body decays, exploiting approximate flavor symmetries in order to cancel the unknown parameters.  
An experimental program for measuring the properties of these decays has been carried out during the last decade at the $B$ factories~\cite{Lees:2013bb,PhysRevD.87.031103} and at the Tevatron~\cite{Aaltonen:2011qt}, and is now continued by LHCb with increased sensitivity. 
The discovery of direct \CP violation in the $B^0 \rightarrow K^+\pi^-$ decay dates back to 2004~\cite{Aubert:2004qm,Chao:2004jy}. This observation raised the question
of whether the effect could be accommodated by the SM or was due to non-SM physics. A simple but powerful model-independent test was proposed in Refs.~\cite{He:1998rq,Lipkin:2005pb}, which required the measurement of direct \CP violation in the $B^0_s \rightarrow K^-\pi^+$ decay. However, \CP violation has never been observed with significance exceeding five Gaussian standard deviations ($\sigma$) in any $B^0_s$ meson decay so far. 

In this Letter we report measurements of direct $C\!P$-violating asymmetries in $B^0 \rightarrow K^+\pi^-$ and $B^0_s \rightarrow K^- \pi^+$ decays using $pp$ collision data, corresponding to an integrated luminosity of $1.0~\mathrm{fb}^{-1}$, collected with the LHCb detector in 2011 at a center-of-mass energy of $7\!$~\tev. 
The present results supersede those given in Ref.~\cite{LHCb-PAPER-2011-029}.
The inclusion of charge-conjugate decay modes is implied except in the asymmetry definitions.
The direct $C\!P$ asymmetry in the $B^0_{(s)}$ decay rate to the final state $f_{(s)}$, with $f=K^+\pi^-$ and $f_s=K^-\pi^+$, is defined as
\begin{equation}
A_{C\!P} \!\!\left(\!B^0_{(s)} \!\! \to \!\! f_{(s)} \! \right ) \!\!=\! \Phi \! \left[\Gamma \! \left(\!\overline{B}^0_{(s)} \!\! \rightarrow \! \!\bar{f}_{(s)} \!\right)\!\!,\,\Gamma \! \left(\!B^0_{(s)} \!\!\rightarrow \!\! f_{(s)}\!\right)\right]\!\!,\label{eq:acp}
\end{equation}
where $\Phi[X,\,Y] = (X-Y)/(X+Y)$ and $\bar{f}_{(s)}$ denotes the charge-conjugate of $f_{(s)}$.

The LHCb detector~\cite{Alves:2008zz} is a single-arm forward spectrometer covering the pseudorapidity range $2<\eta<5$, designed for the study of particles containing $b$ or $c$ quarks. 
The trigger~\cite{LHCb-DP-2012-004} consists of a hardware stage, based on information from the calorimeter and muon systems, followed by a software stage that applies a full event reconstruction.
The hadronic hardware trigger selects large transverse energy clusters in the hadronic calorimeter.
The software trigger requires a two-, three-, or four-track secondary vertex with a large sum of the transverse momenta ($p_\mathrm{T}$) of
the tracks and a significant displacement from the primary $pp$ interaction vertices (PVs). At least one track should have $p_\mathrm{T}$ and impact parameter (IP) $\chi^2$ with respect to all PVs exceeding given thresholds. The IP is defined as the distance between the reconstructed trajectory of a particle and a given $pp$ collision vertex, and the IP $\chi^2$ is the difference between the $\chi^2$ of the PV reconstructed with and without the considered track.
A multivariate algorithm is used for the identification of secondary vertices consistent with the decay of a $b$ hadron.
In order to improve the trigger efficiency on hadronic two-body decays, a dedicated two-body software trigger is also used.
This trigger imposes requirements on the following quantities: the quality of the online-reconstructed tracks, their $p_\mathrm{T}$ and IP; the distance of closest approach of the decay products of the $B$ meson candidate, its $p_\mathrm{T}$, IP and the decay time in its rest frame.

\begin{figure*}[t]
\begin{center}
\includegraphics[width=0.77\textwidth]{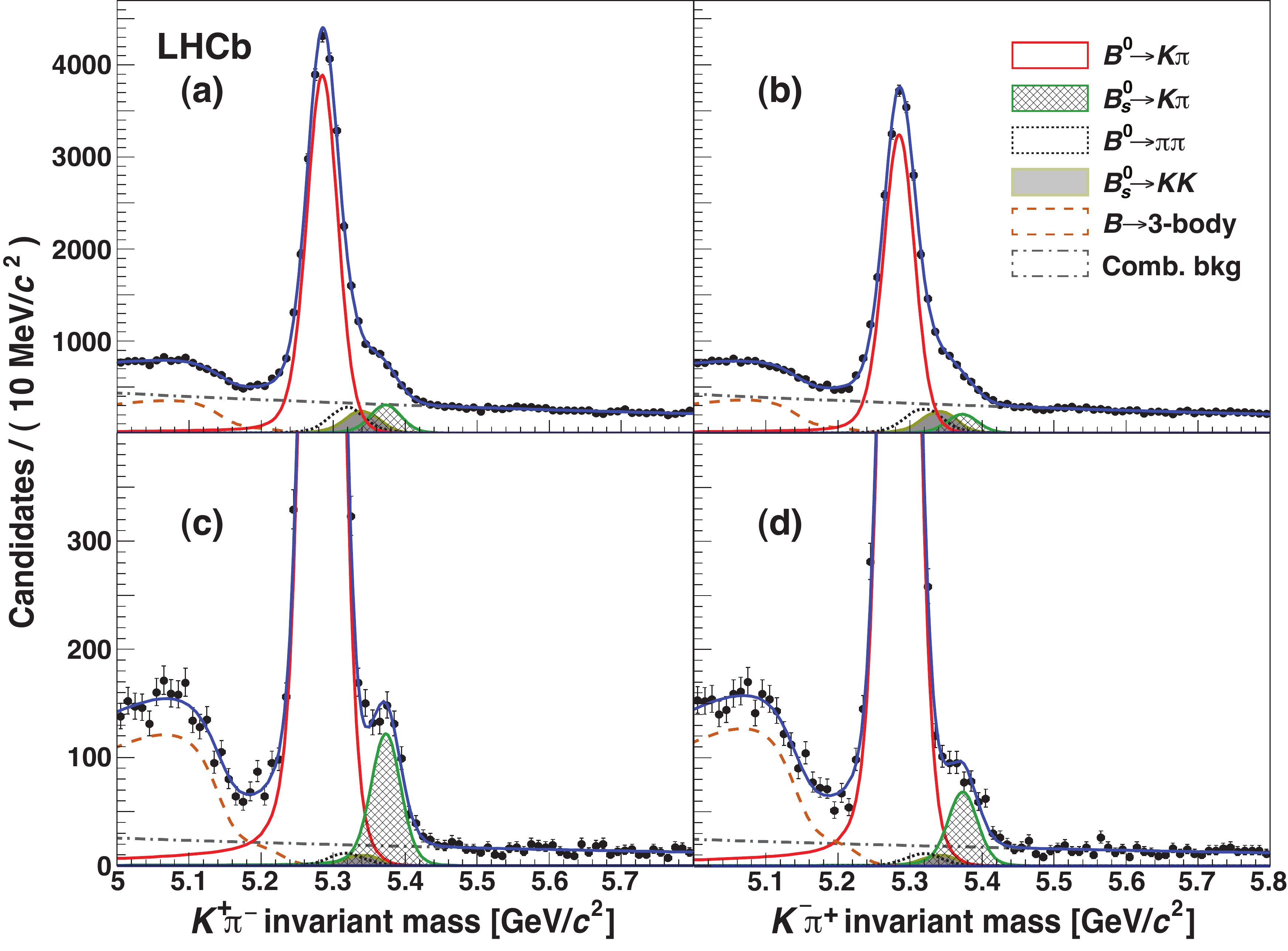} 
\vspace{-0.5cm}
\end{center}
\caption{Invariant mass spectra obtained using the event selection adopted for the best sensitivity on (a, b) $A_{C\!P}(B^0 \rightarrow K^+\pi^-)$ and (c, d) $A_{C\!P}(B^0_s \rightarrow K^-\pi^+)$. Panels (a) and (c) represent the $K^+\pi^-$ invariant mass, whereas panels (b) and (d) represent the $K^-\pi^+$ invariant mass. The results of the unbinned maximum likelihood fits are overlaid. The main components contributing to the fit model are also shown.}
\label{fig:bd2kpifit_bd2kpi2}
\end{figure*}

More selective requirements are applied offline. Two sets of criteria have been optimized with the aim of minimizing the expected statistical uncertainty either on $A_{C\!P}(B^0 \rightarrow K^+\pi^-)$ or on $A_{C\!P}(B^0_s \rightarrow K^- \pi^+)$. In addition to the requirements on the kinematic variables already used in the trigger, requirements on the largest $p_\mathrm{T}$ and IP of the $B$ daughter particles are applied. 
In the case of $B^0_s \rightarrow K^-\pi^+$ decays, a tighter selection is needed to achieve stronger rejection of combinatorial background. For example, the decay time is required to exceed $1.5$~ps, whereas in the $B^0 \rightarrow K^+\pi^-$ selection a lower threshold of $0.9$~ps is applied. This is
because the probability for a $b$ quark to form a $B^0_s$ meson, which subsequently decays to the $K^-\pi^+$ final state, is one order of magnitude smaller than that to form a $B^0$ meson decaying to $K^+\pi^-$~\cite{LHCb-PAPER-2012-002}.
The two samples are then subdivided according to the various final states 
using the particle identification (PID) provided by the two ring-imaging Cherenkov (RICH) detectors~\cite{LHCb-DP-2012-003}.
Two sets of PID selection criteria are applied: a loose set optimized for the measurement of $A_{C\!P}(B^0 \rightarrow K^+\pi^-)$ and a tight set for that of $A_{C\!P}(B^0_s \rightarrow K^-\pi^+)$.
More details on the event selection can be found in Ref.~\cite{LHCb-PAPER-2011-029}.

To determine the amount of background events from other two-body $b$-hadron decays with a misidentified pion or kaon (cross-feed background), the relative efficiencies of the RICH PID selection criteria must be determined. This is achieved by means of a data-driven method that uses $D^{*+} \rightarrow D^0(K^-\pi^+) \pi^+$ and $\L \rightarrow p \pi^-$ decays as control samples.
The production and decay kinematic properties of the $D^0 \rightarrow K^-\pi^+$ and $\L \rightarrow p \pi^-$ channels differ from those of the $b$-hadron decays under study. Since the RICH PID information is momentum dependent, a calibration procedure is performed by reweighting the distributions of the PID variables obtained from the calibration samples, in order to match the momentum distributions of signal final-state particles observed in data.

Unbinned maximum likelihood fits to the mass spectra of the selected events are performed. 
The $B^0 \rightarrow K^+\pi^-$ and $B^0_s \rightarrow K^-\pi^+$ signal components are described by double Gaussian functions convolved with a function that describes the effect of final-state radiation~\cite{Baracchini:2005wp}. The background due to partially reconstructed three-body $B$ decays is parameterized by means of two ARGUS functions~\cite{Albrecht:1989ga} convolved with a Gaussian resolution function. The combinatorial background is modeled by an exponential function and the shapes of the cross-feed backgrounds, mainly due to $B^0 \rightarrow \pi^+\pi^-$ and $B^0_s \rightarrow K^+K^-$ decays with one misidentified particle in the final state, are obtained from simulation. The cross-feed background yields are determined from the $\pi^+\pi^-$, $K^+K^-$, $p\pi^-$ and $pK^-$ mass spectra, using events passing the same selection as the signal and taking into account the appropriate PID efficiency factors. The $K^+\pi^-$ and $K^-\pi^+$ mass spectra for the events passing the two selections are shown in Fig.~\ref{fig:bd2kpifit_bd2kpi2}. The average invariant mass resolution is about $22~\mevcc$.

From the two mass fits we determine the signal yields $N(B^0 \rightarrow K^+\pi^-)=41\hspace{0.5mm}420\pm 300$ and $N(B^0_s \rightarrow K^-\pi^+)=1065 \pm 55$, as well as the raw asymmetries $A_\mathrm{raw}(B^0 \rightarrow K^+\pi^-)=-0.091 \pm 0.006$ and $A_\mathrm{raw}(B^0_{s} \rightarrow K^-\pi^+)=0.28 \pm 0.04$, where the uncertainties are statistical only. In order to derive the $C\!P$ asymmetries from the observed raw asymmetries, effects induced by the detector acceptance and event reconstruction, as well as due to interactions of final-state particles with
the detector material, must be accounted for. Furthermore, the possible presence of a $B^0_{(s)}-\overline{B}^0_{(s)}$ production asymmetry must also be considered.

The $C\!P$ asymmetry is related to the raw asymmetry by $A_{C\!P}=A_\mathrm{raw} - A_\Delta$, where the correction $A_\Delta$ is defined as
\begin{equation}
A_\Delta (B^0_{(s)} \rightarrow K\pi) =\zeta_{d(s)}A_\mathrm{D}(K\pi)+\kappa_{d(s)} A_\mathrm{P}(B^0_{(s)}),
\end{equation}
with $\zeta_{d}=1$ and $\zeta_{s}=-1$.
The instrumental asymmetry $A_\mathrm{D}(K\pi)$ is given in terms of the detection efficiencies $\varepsilon_\mathrm{D}$ of the charge-conjugate final states by
$A_\mathrm{D}(K\pi) = \Phi[\varepsilon_\mathrm{D}(K^-\pi^+),\,\varepsilon_\mathrm{D}(K^+\pi^-)]$, and the production asymmetry $A_\mathrm{P}(B^0_{(s)})$ is defined in terms of the $\overline{B}^0_{(s)}$ and $B^0_{(s)}$ production rates, $R(\overline{B}^0_{(s)})$ and $R(B^0_{(s)})$, as $A_\mathrm{P}(B^0_{(s)}) = \Phi[R(\overline{B}^0_{(s)}),\,R(B^0_{(s)})]$.
The factors $\kappa_d$ and $\kappa_s$ take into account dilutions due to $B^0$ and $B^0_{s}$ meson mixing, respectively.
Their values also depend on event reconstruction and selection, and are $\kappa_d = 0.303 \pm 0.005$ and $\kappa_s = -0.033 \pm 0.003$~\cite{LHCb-PAPER-2011-029}. The factor $\kappa_s$ is ten times smaller than $\kappa_d$, owing to the large $B^0_s$ oscillation frequency.

The instrumental charge asymmetry $A_\mathrm{D}(K\pi)$ is measured from data using
$D^{*+} \rightarrow D^0(K^-\pi^+)\pi^+$ and $D^{*+} \rightarrow D^0(K^-K^+)\pi^+$ decays.
The combination of the time-integrated raw asymmetries of these two decay modes is used to disentangle the various contributions to each raw asymmetry. The presence
of open charm production asymmetries arising from the primary $pp$ interaction constitutes an additional complication.
We write the following equations relating the observed raw asymmetries to the physical \CP asymmetries
\begin{eqnarray}
A^*_\mathrm{raw}(K\pi)&=&A^*_\mathrm{D}(\pi_{s})+A^*_\mathrm{D}(K\pi)+A_\mathrm{P}(D^{*})\label{eq:acp_raw_dstar_kpi},\\[3mm]
A^*_\mathrm{raw}(KK)&=&A_{C\!P}(KK)+A^*_\mathrm{D}(\pi_{s})+A_\mathrm{P}(D^{*})\label{eq:acp_raw_dstar_kk},
\end{eqnarray} 
where $A^*_\mathrm{raw}(K\pi)$ and $A^*_\mathrm{raw}(KK)$ are the time-integrated raw asymmetries in $D^*$-tagged $D^0 \to K^-\pi^+$ and $D^0 \to K^-K^+$ decays, respectively; $A_{C\!P}(KK)$ is the $D^0 \to K^-K^+$ \CP asymmetry; $A^*_\mathrm{D}(K\pi)$ is the detection asymmetry in reconstructing $D^0 \to K^{-}\pi^{+}$ and $\overline{D}^0 \to K^{+}\pi^{-}$ decays; $A^*_\mathrm{D}(\pi_{s})$ is the detection asymmetry in reconstructing positively- and negatively-charged pions originated from $D^*$ decays; and $A_\mathrm{P}(D^{*})$ is the production asymmetry for prompt charged $D^{*}$ mesons. In Eq.~(\ref{eq:acp_raw_dstar_kpi}) any possible \CP asymmetry in the Cabibbo-favored $D^0 \to K^-\pi^+$ decay is neglected~\cite{Bianco:2003vb}.
By subtracting Eqs.~(\ref{eq:acp_raw_dstar_kpi}) and~(\ref{eq:acp_raw_dstar_kk}), one obtains
\begin{equation}
A^*_\mathrm{raw}(K\pi) - A^*_\mathrm{raw}(KK) = A^*_\mathrm{D}(K\pi) - A_{C\!P}(KK) \label{eq:acp_raw_sub}.
\end{equation}
Once the raw asymmetries are measured, this equation determines unambiguously the detection asymmetry $A^*_\mathrm{D}(K\pi)$, using the world average for the \CP asymmetry of the $D^0 \to K^-K^+$ decay. Since the measured value of the time-integrated asymmetry depends on the decay-time acceptance, the existing measurements of $A_{C\!P}(KK)$~\cite{Staric:2008rx,Aubert:2007if,Aaltonen:2011se} are corrected for the difference in acceptance with respect to LHCb~\cite{LHCb-PAPER-2011-023}. This leads to the value $A_{C\!P}(KK)=(-0.24 \pm 0.18)\%$.
Furthermore, $B$ meson production and decay kinematic properties differ from those of the $D$ decays being considered, and different trigger and selection algorithms are applied. In order to correct the raw asymmetries of $B$ decays, using the detection asymmetry $A^*_\mathrm{D}(K\pi)$ derived from $D$ decays, a reweighting procedure is needed. We reweight the $D^0$ momentum, transverse momentum and azimuthal angle in $D^0 \to K^-\pi^+$ and $D^0 \to K^-K^+$ decays, to match the respective $B^0_{(s)}$ distributions in $B^0 \to K^+\pi^-$ and $B^0_s \to K^-\pi^+$ decays.
The raw asymmetries are determined by means of $\chi^2$ fits to the reweighted $\delta m = M_{D^*} - M_{D^0}$ distributions, where $M_{D^*}$ and
$M_{D^0}$ are the reconstructed $D^*$ and $D^0$ candidate invariant masses, respectively.

From the raw asymmetries, values for the quantity $\Delta A = A_\mathrm{D}(K\pi) - A_{C\!P}(KK)$ are determined.
We obtain the values $\Delta A = (-0.91 \pm 0.15)\%$ and $\Delta A = (-0.98 \pm 0.11)\%$, using as target kinematic distributions those of $B$ candidates passing the event selection optimized for $A_{C\!P}(B^0 \to K^+\pi^-)$ and for $A_{C\!P}(B^0_s \to K^-\pi^+)$, respectively.
Using these two values of $\Delta A$ and the value of $A_{C\!P}(KK)$, we obtain the instrumental asymmetries $A_\mathrm{D}(K\pi)=(-1.15 \pm 0.23)\%$ for the $B^0 \to K^+\pi^-$ decay and $A_\mathrm{D}(K\pi)=(-1.22 \pm 0.21)\%$ for the $B^0_s \to K^-\pi^+$ decay.

Assuming negligible \CP violation in the mixing, as expected in the SM and confirmed by current experimental determinations~\cite{bib:hfagbase}, the decay rate of a $B^0_{(s)}$ meson with production asymmetry $A_\mathrm{P}$, decaying into a flavor-specific final state $f_{(s)}$ with \CP asymmetry $A_{C\!P}$ and detection asymmetry $A_\mathrm{D}$, can be written as
\begin{equation}\label{eq:decayrate}
\mathcal{R}(t;\,p) \propto  \left(1\!-\!p A_{\CP}\right)\left(1\!-\!p A_\mathrm{D}\right)\left[H_{+}\!\left(t\right)\!-\!p A_\mathrm{P}H_{-}\!\left(t\right)\right]\!,
\end{equation}
where $t$ is the reconstructed decay time of the \B meson and $p$ assumes the values $p = +1$ for the final state $f_{(s)}$ and $p = -1$ for the final state $\bar{f}_{(s)}$.
The functions $H_{+}\left(t\right)$ and $H_{-}\left(t\right)$ are defined as
\begin{eqnarray}
  H_{+}\!\left(t\right) \!&=&\!\!
  \left[e^{-\Gamma_{d(s)} t^{\prime}}\!\!\!\cosh\!{\left(\!\!\frac{\Delta\Gamma_{d(s)}}{2}t^{\prime}\!\!\right)}\!\!\otimes \! R\!\left(t,\,t^{\prime}\right)\right]\!\varepsilon_{d(s)}\!\left(t\right)\!, \\
  H_{-}\!\left(t\right) \!&=& \!\!
  \left[e^{-\Gamma_{d(s)} t^{\prime}}\!\!\!\cos\!{\left(\Delta m_{d(s)} t^{\prime}\right)}\!\otimes \!R\!\left(t,\,t^{\prime}\right)\right]\!\varepsilon_{d(s)}\!\left(t\right)\!, 
\end{eqnarray}
where $\Gamma_{d(s)}$ is the average decay width of the $B^0_{(s)}$ meson, $\Delta\Gamma_{d(s)}$ and $\Delta m_{d(s)}$ are the decay width and mass differences between the two $B^0_{(s)}$ mass eigenstates respectively, $R\left(t,\,t^{\prime}\right)$ is the decay time resolution ($\sigma \simeq 50~\mathrm{fs}$ in our case) and the symbol $\otimes$ stands for convolution. Finally $\varepsilon_{d(s)}\left(t\right)$ is the acceptance as a function of the $B^0_{(s)}$ decay time.
Using Eq.~(\ref{eq:decayrate}) we obtain the following expression for the time-dependent asymmetry
\begin{eqnarray}
   \mathcal{A}\left(t\right) & = & \Phi \! \left[ \mathcal{R}\left(t;\,-1\right) \!,\, \mathcal{R}\left(t;\,+1\right) \right]  \label{eq:timeAsymmetry} \\
    & = & \frac{\left(A_{\CP}\!+\!A_\mathrm{D}\right)\!H_{+}\!\left(t\right)\!+\!A_\mathrm{P}\left(1\!+\!A_{\CP}A_\mathrm{D}\right)\!H_{-}\!\left(t\right)}{\left(1\!+\!A_{\CP}A_\mathrm{D}\right)\!H_{+}\!\left(t\right)\!+\!A_\mathrm{P}\left(A_{\CP}\!+\!A_\mathrm{D}\right)\!H_{-}\!\left(t\right)}.\nonumber 
\end{eqnarray}
For illustrative purposes only, we consider the case of perfect decay time resolution and negligible $\Delta \Gamma$, retaining only first-order terms in $A_{C\!P}$, $A_\mathrm{P}$ and $A_\mathrm{D}$. In this case, Eq.~(\ref{eq:timeAsymmetry}) reduces to the expression
\begin{equation}\label{eq:timeAsymmetrySimple}
   \mathcal{A}\left(t\right) \approx A_{\CP}+A_\mathrm{D}+A_\mathrm{P}\cos{\left(\Delta m_{d(s)}t\right)},
\end{equation}
{\it i.e.}, the time-dependent asymmetry has an oscillatory term with amplitude equal to the production asymmetry $A_\mathrm{P}$. By studying the full time-dependent decay rate it is then possible to determine $A_\mathrm{P}$ unambiguously.

\begin{figure}[t]
  \begin{center}
    \includegraphics[width=0.48\textwidth]{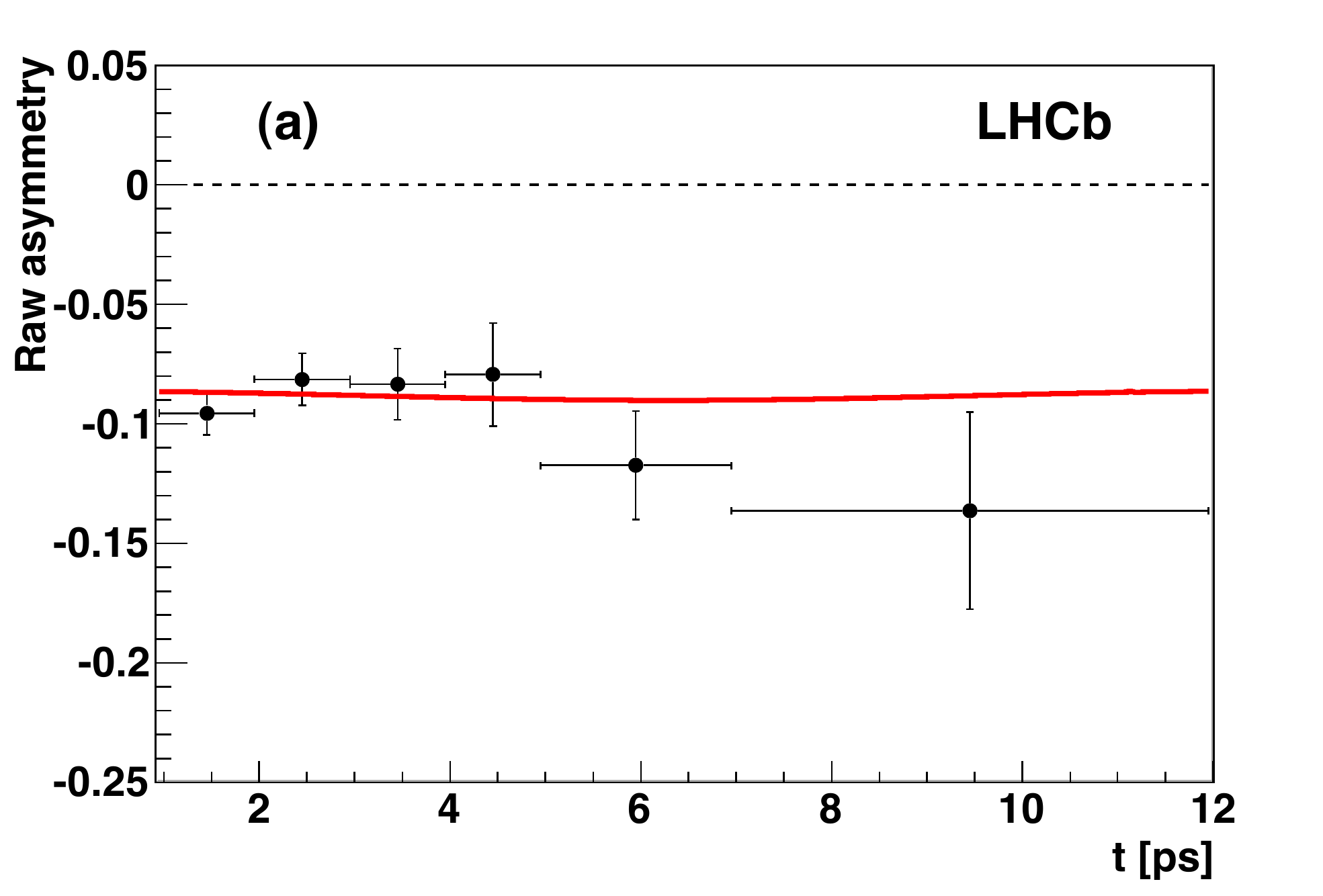}
    \includegraphics[width=0.48\textwidth]{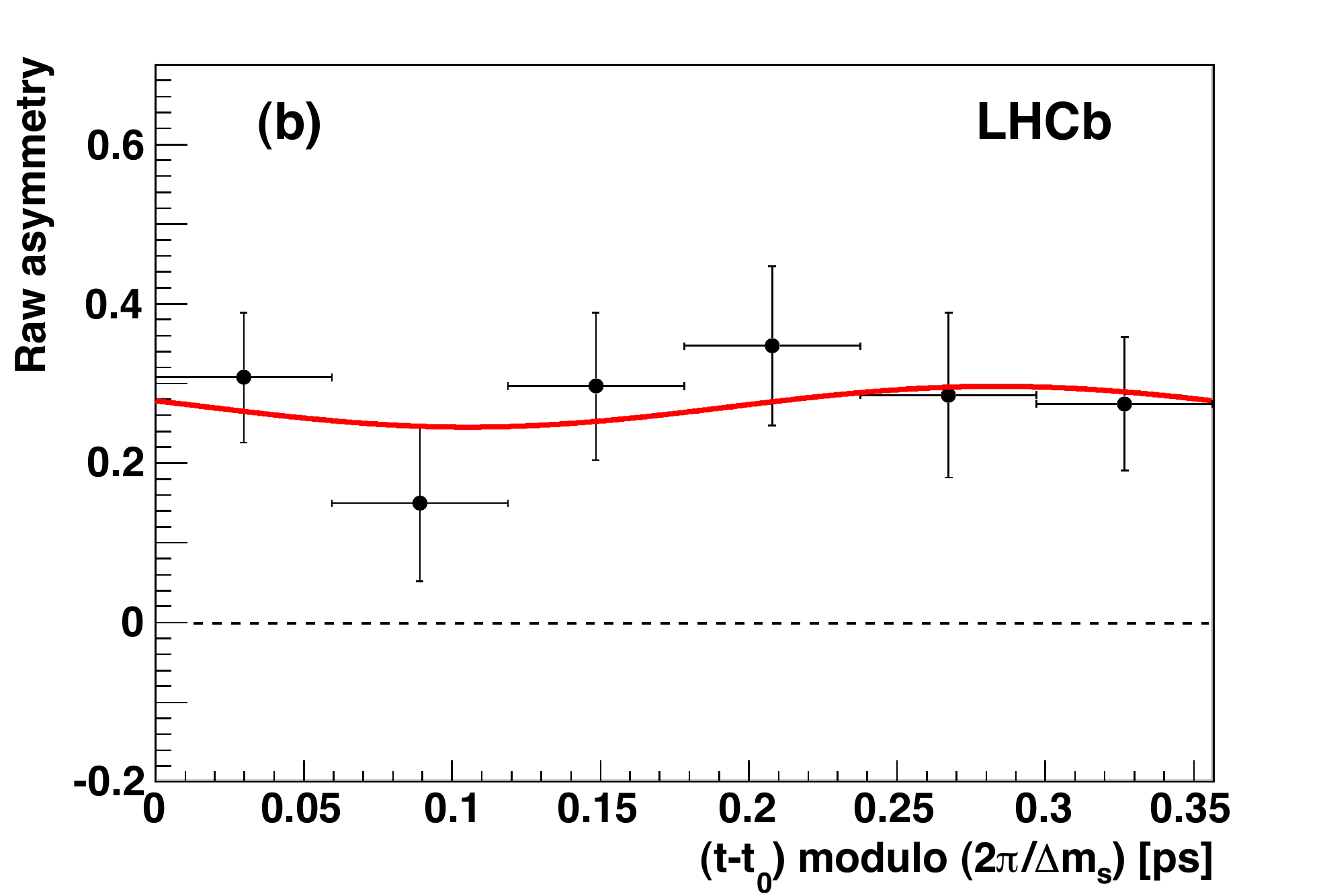}
 \end{center}
  \caption{Raw asymmetries as a function of the decay time for (a) $B^0 \to K^+\pi^-$ and (b) $B^0_s \to K^-\pi^+$ decays. In (b), the offset $t_0=1.5$\ps corresponds to the minimum value of the decay time required by the $B^0_s \to K^-\pi^+$ event selection. The curves represent the asymmetry projections of fits to the decay time spectra.}
  \label{fig:apAsym}
\end{figure}

In order to measure the production asymmetry $A_\mathrm{P}$ for \Bd and \Bs mesons, we perform fits to the decay time spectra of the \B candidates, separately for the events passing the two selections. The \Bd production asymmetry is determined from the sample obtained applying the selection optimized for the measurement of $A_{\CP}(B^0 \to K^+\pi^-)$, whereas the \Bs production asymmetry is determined from the sample obtained applying the selection optimized for the measurement of $A_{\CP}(B^0_s \to K^-\pi^+)$. We obtain $A_\mathrm{P}(B^0) = (0.1 \pm 1.0)\%$  and $A_\mathrm{P}(B^0_s) = (4 \pm 8)\%$. Figure~\ref{fig:apAsym} shows the raw asymmetries as a function of the decay time, obtained by performing fits to the invariant mass distributions of events restricted to independent intervals of the $B$ candidate decay times.

By using the values of the detection and production asymmetries, the correction factors to the raw asymmetries $A_\Delta (B^{0}\rightarrow K^+\pi^-) = \left (-1.12 \pm 0.23 \pm 0.30 \right)\!\%$ and
$A_\Delta (B^{0}_s\rightarrow K^- \pi^+) = \left( 1.09 \pm 0.21 \pm 0.26 \right)\!\%$ are obtained, where the first uncertainties are due to the detection asymmetry and the second to the production asymmetry.

Systematic uncertainties on the asymmetries are related to PID calibration, modeling of the signal and background components in the maximum likelihood fits and instrumental charge asymmetries. 
In order to estimate the impact of imperfect PID calibration, we perform mass fits to determine raw asymmetries using altered numbers of cross-feed background events, according to the systematic uncertainties affecting the PID efficiencies.
An estimate of the uncertainty due to possible mismodeling of the final-state radiation is determined by varying the amount of emitted
radiation~\cite{Baracchini:2005wp} in the signal shape parameterization, according to studies performed on fully simulated events, in which final state radiation is generated using \photos~\cite{Golonka:2005pn}.  The possibility of an incorrect description of the signal mass model is investigated by replacing the double Gaussian function with the sum of three Gaussian functions, where the third component has fixed fraction ($5\%$) and width ($50\,\mevcc$), and is aimed at describing long tails, as observed in simulation. 
To assess a systematic uncertainty on the shape of the partially reconstructed backgrounds, we remove the second ARGUS function.
For the modeling of the combinatorial background component, the fit is repeated using a straight line. Finally, for the case of the cross-feed backgrounds, two distinct systematic uncertainties are estimated: one due to a relative bias in the mass scale of the simulated distributions with respect to the signal distributions in data, and another accounting for the difference in mass resolution between simulation and data. All shifts from the relevant baseline values are accounted for as systematic uncertainties.
Systematic uncertainties related to the determination of detection asymmetries are calculated by summing in quadrature the respective uncertainties on $A_\Delta (B^{0}\rightarrow K^+\pi^-)$ and $A_\Delta (B^{0}_s\rightarrow K^- \pi^+)$ with an additional uncertainty of $0.10\%$,  
accounting for residual differences in the trigger composition between signal and calibration samples.

The systematic uncertainties for $A_{C\!P}(B^0 \rightarrow K^+ \pi^-)$ and $A_{C\!P}(B^0_s \rightarrow K^-\pi^+)$ are summarized in Table~\ref{tab:SystsumB}.
Since the production asymmetries are obtained from the fitted decay time spectra of $B^0 \to K^+\pi^-$ and $B^0_s \to K^-\pi^+$ decays, their uncertainties are statistical in nature and are then propagated to the statistical uncertainties on $A_{C\!P}(B^0 \rightarrow K^+ \pi^-)$ and $A_{C\!P}(B^0_s \rightarrow K^-\pi^+)$.

  \begin{table}
    \caption{Systematic uncertainties on $A_{C\!P}(B^0 \rightarrow K^+ \pi^-)$ and $A_{C\!P}(B^0_s \rightarrow K^-\pi^+)$. The total systematic uncertainties are obtained by summing the individual contributions in quadrature.}\label{tab:SystsumB}
    \begin{center}
\resizebox{0.48\textwidth}{!}{
      \begin{tabular}{lcc}
        \hline
\hline
Systematic uncertainty  & $A_{C\!P}(B^0 \rightarrow K^+ \pi^-)$ \bigstrut & $A_{C\!P}(B^0_s \rightarrow K^-\pi^+)$ \\
\hline
PID calibration                                           & 0.0006 & 0.0012 \\
Final state radiation                                   & 0.0008 & 0.0020 \\
Signal model                                              & 0.0001 & 0.0064 \\
Combinatorial background             & 0.0004 & 0.0042 \\
Three-body background                       & 0.0005 & 0.0027 \\
Cross-feed background        & 0.0010 & 0.0033 \\
Detection asymmetry          & 0.0025   & 0.0023  \\
\hline 
Total                                                           & 0.0029  & 0.0094 \\
\hline\hline
      \end{tabular}
}
    \end{center}
\vspace{-0.3cm}
  \end{table}

In conclusion, the parameters of $C\!P$ violation in $B^0 \to K^+\pi^-$ and $B^0_s \to K^-\pi^+$ decays have been measured to be
\begin{eqnarray}
A_{C\!P}(B^0 \!\rightarrow \! K^+\pi^-)&=&-0.080 \pm 0.007\,\mathrm{(stat)} \pm 0.003\,\mathrm{(syst)},\nonumber \\[3mm]
A_{C\!P}(B^0_s \!\rightarrow \! K^-\pi^+)&=&0.27 \pm 0.04\, \mathrm{(stat)}\pm 0.01\,\mathrm{(syst)}.\nonumber
\end{eqnarray}
Dividing the central values by the sum in quadrature of statistical and systematic uncertainties, the significances of the measured deviations from zero are $10.5\sigma$ and $6.5\sigma$, respectively. The former is the most precise measurement of $A_{C\!P}(B^0 \to K^+\pi^-)$ to date, whereas the latter represents the first observation of \CP violation in decays of $B^0_s$ mesons with significance exceeding $5\sigma$.
Both measurements are in good agreement with world averages~\cite{bib:hfagbase} and previous LHCb results~\cite{LHCb-PAPER-2011-029}.

These results allow a stringent test of the validity of the relation between $A_{C\!P}(B^0 \to K^+\pi^-)$ and $A_{C\!P}(B^0_s \rightarrow K^-\pi^+)$ in the SM given in Ref.~\cite{Lipkin:2005pb} as
\begin{equation}
\Delta = \frac{A_{C\!P}(B^0 \!\rightarrow \! K^+\pi^-)}{A_{C\!P}(B^0_s \!\rightarrow \! K^-\pi^+)} + \frac{\mathcal{B}(B^0_s \!\to \! K^-\pi^+)}{\mathcal{B}(B^0 \!\to \! K^+\pi^-)} \frac{\tau_d}{\tau_s} = 0,\label{lipkincheck}
\end{equation}
where $\mathcal{B}(B^0 \to K^+\pi^-)$ and $\mathcal{B}(B^0_s \to K^-\pi^+)$ are $C\!P$-averaged branching fractions, and $\tau_d$ and $\tau_s$ are the $B^0$ and $B^0_s$ mean lifetimes, respectively. Using additional results for $\mathcal{B}(B^0 \to K^+\pi^-)$ and $\mathcal{B}(B^0_s \to K^-\pi^+)$~\cite{LHCb-PAPER-2012-002} and the world averages for $\tau_{d}$ and $\tau_{s}$~\cite{bib:hfagbase}, we obtain $\Delta = -0.02 \pm 0.05 \pm 0.04$, where the first uncertainty is from the measurements of the $C\!P$ asymmetries and the second is from the input values of the branching fractions and the lifetimes. No evidence for a deviation from zero of $\Delta$ is observed with the present experimental precision.

\section*{Acknowledgements}

\noindent We express our gratitude to our colleagues in the CERN
accelerator departments for the excellent performance of the LHC. We
thank the technical and administrative staff at the LHCb
institutes. We acknowledge support from CERN and from the national
agencies: CAPES, CNPq, FAPERJ and FINEP (Brazil); NSFC (China);
CNRS/IN2P3 and Region Auvergne (France); BMBF, DFG, HGF and MPG
(Germany); SFI (Ireland); INFN (Italy); FOM and NWO (The Netherlands);
SCSR (Poland); ANCS/IFA (Romania); MinES, Rosatom, RFBR and NRC
``Kurchatov Institute'' (Russia); MinECo, XuntaGal and GENCAT (Spain);
SNSF and SER (Switzerland); NAS Ukraine (Ukraine); STFC (United
Kingdom); NSF (USA). We also acknowledge the support received from the
ERC under FP7. The Tier1 computing centres are supported by IN2P3
(France), KIT and BMBF (Germany), INFN (Italy), NWO and SURF (The
Netherlands), PIC (Spain), GridPP (United Kingdom). We are thankful
for the computing resources put at our disposal by Yandex LLC
(Russia), as well as to the communities behind the multiple open
source software packages that we depend on.

\addcontentsline{toc}{section}{References}
\bibliographystyle{LHCb}
\bibliography{main,paper-aps,LHCb-PAPER,LHCb-CONF,LHCb-DP}

\end{document}